\documentclass[fleqn,twoside]{article}

\usepackage{amsmath}
\usepackage{cite}
\input{xy}
\xyoption{all}

\newcommand{\pisiSE}{$\Pi\Sigma^*$}

\makeatletter
 \def\@problemhead#1#2#3{%
  \par\kern-\parskip\kern#1
  \vbox\bgroup
  \hbox to\hsize{\hrulefill\raise1pt\hbox{\fbox{$\mathstrut$ #3\ }}\hrulefill}%
  \kern#2\par\kern-\parskip
 }
 \def\@problemtail#1#2{%
  \par\kern-\parskip\kern#1
  \hbox to\hsize{\hrulefill}
  \egroup
  \kern#2\par\kern-\parskip
 }
\newenvironment{ProblemSpec}[1]{\@problemhead{2pt}{0pt}{#1}\small}{\@problemtail{-8pt}{2pt}}
\makeatother

\newcommand{\SigmaP}{\texttt{Sigma}}
\newcommand{\HarP}{\texttt{HarmonicSums}}

\usepackage{espcrc2}

\usepackage{graphicx}

\title{
\vspace*{-5mm} \noindent
{\tiny \begin{flushleft} DESY 10--90 \hfill TTK 10--37 \hfill
                                SFB/CPP-10-53   \\
              \end{flushleft}}
Modern Summation Methods and the Computation of 2- and 3-loop Feynman Diagrams}

\author{Jakob Ablinger~\address[RISC]{Research Institute for Symbolic Computation (RISC)\\ Johannes
Kepler University Linz, Altenberger Str. 69, A--4040 Linz, Austria}~\thanks{Supported by the Austrian
Science Fund (FWF) grant P20347-N18.}, Johannes Bl\"umlein~\address[DESY]{Deutsches
Elektronen-Synchrotron (DESY), Platanenallee 6, D--15738 Zeuthen, Germany}~\thanks{Supported in part by
SFB-TR-9
and EU TMR network HEPTOOLS},
        Sebastian Klein~\address{
Institut f\"ur Theoretische Teilchenphysik und Kosmologie, RWTH Aachen University, D--52056 Aachen,
Germany}~\thanks{Supported in part by SFB-TR-9},
Carsten Schneider~\addressmark[RISC]\thanks{Supported by the Austrian Science Fund (FWF) grants
P20162-N18 and P20347-N18.}}

\begin{document}

\begin{abstract}
By symbolic summation methods based on difference fields we present a general
strategy that transforms definite multi-sums, e.g., in terms of hypergeometric
terms and harmonic sums, to indefinite nested sums and products. We succeeded
in this task with all our concrete calculations of 2--loop and 3--loop massive
single scale Feynman diagrams with local operator insertion.
\vspace{-1pc}
\end{abstract}

\maketitle

\section{Introduction}

We aim at the simplification of multi--sums to expressions in terms of indefinite nested sums and
products. The calculations are performed in the context of the computation of 2- and 3-loop massive
single scale Feynman diagrams with local operator insertion. These are related to the QCD anomalous
dimensions and massive operator matrix elements. The mathematical expressions depend on the Mellin variable $N$, defined either at the even or odd positive integers. The Feynman diagrams may be mapped to definite nested sums over hypergeometric expressions involving also harmonic sums~\cite{Vermaseren:99,Bluemlein:99}. Given these usually huge expressions, the task is to find simpler representations that can be processed further in physics.
At 2-loop order all respective calculations are finished~\cite{Neerven:96,BBK:07,BBKS:08}
and lead to representations in terms of harmonic sums only.
This is not necessarily the case from 3-loop order onwards. In particular, during our calculations, we left the solution space of harmonic sums, and only in the end, while combining our results, we end up again at harmonic sums or their generalized versions~\cite{Moch:02}.

Inspired by symbolic summation of hypergeometric terms~\cite{AequalB}, in particular by creative telescoping~\cite{Zeilberger:91} and finding hypergeometric solutions of linear recurrences~\cite{Petkov:92}, we developed a general summation machinery for indefinite nested sums and products. Based on a refined version~\cite{Schneider:08c} of Karr's summation theory of \pisiSE--difference fields~\cite{Karr:81} we use algorithms from~\cite{Schneider:05c,Schneider:05a,Schneider:08c,Schneider:10c,Schneider:10b}  implemented in the Mathematica package \SigmaP~\cite{Schneider:07a} that attacks the summation quantifiers from the innermost sums to the outermost sums and which transforms step by step the sums under consideration to indefinite nested sums and products. Finally, we use the
Mathematica package~\HarP~\cite{Ablinger:09} inspired by various ideas
of~\cite{Vermaseren:99,Bluemlein:99,Moch:02,Bluemlein:04,Blumlein:2009ta} and extensions of it~\cite{ABS:10} in order to transform the indefinite nested product-sum expressions --whenever possible-- to harmonic sums or its generalized versions~\cite{Moch:02}. In particular, combining the results, the elimination of algebraic relations among the occurring harmonic sums plays a crucial role.

For the physical background and further results in our computations we refer to~\cite{LoopLegPhysics}.

\section{Simplification of multisums}

We consider the following problem: \textit{Given} a definite nested multi-sum over hypergeometric terms\footnote{$f(j)$ is hypergeometric in $j$ iff $f(j+1)/f(j)=r(j)$ for some fixed rational function $r(j)$.} involving also harmonic sums, for instance
\begin{multline}\label{Equ:OrgSumRunningExp}
F(N)=\sum_{j=0}^{N-2} \sum_{r=0}^{j+1} \sum_{s=0}^{-j+N+r-2}\\ \frac{(-1)^{r+s} \
\binom{j+1}{r} \binom{-j+N+r-2}{s} (-j+N-2)! r!}{(N-s) (s+1) (-j+N+r)!},
\end{multline}
\textit{find} an alternative representation in terms of indefinite nested product--sum expressions. This means in particular indefinite nested sums over hypergeometric terms (like binomials, factorials, Pochhammer symbols) that may occur as polynomial expressions in the numerator and denominator with the additional constraint that the summation index $i_j$ of a sum $\sum_{i_j=1}^{i_{j+1}}f(i_j)$ may occur only as the upper index of its inner sums and products, but not inside of the inner sums itself; for a precise but rather technical definition we refer to~\cite{Schneider:10b}. Typical examples are harmonic sums~\cite{Vermaseren:99,Bluemlein:99} defined by
\begin{multline}\label{Equ:HarmonicSums}
S_{m_1,\dots,m_k}(N)=\\
\sum_{i_1=1}^N\frac{\text{\small$\text{sign}(m_1)^{i_1}$}}{i_1^{|m_1|}}\dots
\sum_{i_k=1}^{i_{k-1}}\frac{\text{\small$\text{sign}(m_k)^{i_k}$}}{i_k^{|m_k|}}
\end{multline}
with $m_1,\dots,m_k$ being nonzero integers, their generalized versions called $S$--sums~\cite{Moch:02}
\begin{multline}\label{Equ:SSums}
S_{m_1,\dots,m_k}(x_1,\dots,x_k,N)=
\sum_{i_1=1}^n\frac{x_1^{i_1}}{i_1^{m_1}}\dots
\sum_{i_k=1}^{i_{k-1}}\frac{x_k^{i_k}}{i_k^{m_k}},
\end{multline}
with constants $x_i$ or, e.g., nested binomial sums of the form
\begin{align}\label{BinomialSums}
\sum_{i=1}^N \frac{\displaystyle\sum_{j=1}^i \frac{\binom{2 j}{j}}{j}}{\binom{2 \
i}{i}},\quad
\sum_{i=1}^N \frac{\displaystyle S_1(i)\sum_{j=1}^i \frac{\binom{2 j}{j}}{j}}{\binom{2 i}{i}};
\end{align}
examples for these sums are given in Section~\ref{Sec:Examples}.
Note that the sum~\eqref{Equ:OrgSumRunningExp} is not given in indefinite nested form since, e.g., the summation index $j$ of the outermost sum occurs inside in the summands of the inner sums; for a transformation to indefinite nested sums see Subsection~\ref{Sec:DetailedExp}.

\subsection{The underlying summation principles}

In order to transform sums such as~\eqref{Equ:OrgSumRunningExp} to indefinite nested sums, we use the summation package \SigmaP~\cite{Schneider:07a} by the following strategy; the underlying algorithms~\cite{Schneider:05c,Schneider:05a,Schneider:08c,Schneider:10c,Schneider:10b} are based on various extensions and refinements of Karr's summation theory of \pisiSE-difference fields~\cite{Karr:81}. We process each summation quantifier and transform step by step the occurring sums to indefinite nested versions. Suppose, e.g., that we derived already an expression $f(N,j)$ in terms of indefinite nested sums and products w.r.t.\ $j$ and that we want to attack the next summation quantifier, say $\sum_{j=0}^{N-2}f(N,j)$. If $f(N,j)$ is free of $N$, we are already done. Otherwise, we proceed by creative telescoping.

\begin{ProblemSpec}{Deriving recurrences by creative telescoping}
\noindent{\it
Given} an integer $d>0$ and given a sum

\vspace*{-0.3cm}

\begin{equation}\label{Equ:CreaSum}
F(a,N):=\sum_{j=0}^af(N,j)
\end{equation}

\vspace*{-0.2cm}

\noindent with an extra parameter $N$, {\it find} constants
$c_0(N),\dots,c_d(N)$, free of $j$, and $g(N,j)$ such that for $0\leq j\leq a$ the
following summand recurrence holds:
\begin{multline}\label{Equ:CreaTele}
c_0(N)f(N,j)+\dots+c_d(N)f(N+d,j)\\
=g(N,j+1)-g(N,j).
\end{multline}
\end{ProblemSpec}

\noindent Creative telescoping has been originally introduced with Zeilberger's algorithm~\cite{Zeilberger:91,Paule:95} where the summand $f(N,j)$ can be a hypergeometric term and the derived $c_i(N)$ are rational functions in $N$ and $g(N,j)$ is a rational multiple of $f(N,j)$. More generally,
with the summation package \SigmaP\ the summand $f(N,j)$ can be an indefinite nested product--sum 
expression w.r.t. $j$, and one searches for constants $c_i(N)$ which are indefinite nested 
product--sum expressions in $N$ and expressions $g(N,j)$ which are indefinite nested product--sum expressions w.r.t.\ $j$.\\
\noindent If one succeeds in
this task to compute a summand recurrence, one gets by telescoping the recurrence relation
\begin{multline}\label{Equ:SummandRec}
c_0(N)F(a,N)+\dots+c_d(N)F(a,N+d)\\
=g(N,a+1)-g(N,0).
\end{multline}

\textit{Remark.} In most instances, $a$ depends linearly on $N$, which means that $g(N,a+1)$ itself might be a definite sum. In this case, one first has to transform $g(N,a+1)$ to an indefinite nested product--sum representation by recursive application of our strategy. In other instances, $a$ has to be sent to infinity to deal with sums of the form $\sum_{j=0}^{\infty}f(N,j)$.

\smallskip

If one succeeds in this task, we obtain
\begin{equation}\label{Equ:RecGeneral}
c_0(N)F(N)+\dots+c_d(N)F(N+d)=h(N)\\
\end{equation}
where the $c_i(N)$ and $h(N)$ are given in terms of indefinite nested product--sums. Then we can apply another feature of \SigmaP.

\begin{ProblemSpec}{Recurrence solving\index{recurrence!solving}}
\noindent{\it Given} a
recurrence of the form \eqref{Equ:RecGeneral}, {\it find} all solutions in terms of indefinite nested product--sum expressions (also called d'Alembertian solution).
\end{ProblemSpec}
\noindent Based on the underlying algorithms, see e.g.~\cite{Abramov:94,BKKS:09}, the derived 
d'Alem\-bertian solutions of~\eqref{Equ:RecGeneral} are highly nested: in the worst case the sums will 
reach the nesting depth $d-1$. In order to simplify these solutions, a refined telescoping paradigm~\cite{Schneider:10b} based on depth--optimal \pisiSE-difference fields~\cite{Schneider:08c}
is activated. This machinery delivers algebraic independent sum representations~~\cite{Schneider:10c} with minimal nesting depth~\cite{Schneider:10c}.

To this end, one tries to combine these solutions such that the evaluation agrees with the original sum $F(N)$ for the first $N=0,...,d-1$ initial values. Then since both expressions are a solution of a recurrence of order $d$, the expressions agree --up to some mild side conditions-- for all $N\geq0$.

As worked out in this subsection, \SigmaP\ 
is a general symbolic summation toolbox that assists in the task 
to transform definite multi-sums to indefinite nested product--sums. 
At this point we should mention that the strategy sketched above might fail: 
it might go wrong to represent $f(N,j)$ in~\eqref{Equ:CreaSum} in terms of indefinite 
nested sums or products, the Ansatz of creative telescoping might fail to find a 
recurrence of the form~\eqref{Equ:RecGeneral}, or we might miss sufficiently many 
solutions (i.e., some of the solutions of~\eqref{Equ:RecGeneral} are not expressible 
in terms of indefinite nested product--sum expressions)  so that they can be combined 
to an alternative representation of the input sum~\eqref{Equ:CreaSum}. However, for all 
examples that we have encountered so far in 2--loop and 3--loop computations, the machinery never failed.

For the transformation of our results to harmonic sums and their generalized versions, J.~Ablinger's 
\HarP\ package~\cite{Ablinger:09} plays an important role. It is inspired by various ideas of~\cite{Vermaseren:99,Bluemlein:99,Moch:02,Bluemlein:04,Blumlein:2009ta} and extensions of it~\cite{ABS:10}; some highlights are given in Section~\ref{Sec:Examples}.

\subsection{A detailed example}\label{Sec:DetailedExp}

Subsequently, we illustrate this interplay of the different \SigmaP-functions for telescoping, creative telescoping and recurrence solving in the difference field setting by the given sum~\eqref{Equ:OrgSumRunningExp} which can be rewritten in the form
\begin{multline}\label{Equ:TripleSum}
\sum_{j=0}^{N-2} (-j+N-2)! \sum_{r=0}^{j+1} \frac{(-1)^r \binom{j+1}{r} r!}{(-j+N+r)!}\times\\
\times\sum_{s=0}^{-j+N+r-2} \frac{(-1)^s \binom{-j+N+r-2}{s}}{(N-s)
(s+1)}.
\end{multline}
We start with the innermost sum
\begin{equation}\label{Equ:InnerSum1}
F_0(N,j,r)=\sum_{s=0}^{-j+N+r-2} \frac{(-1)^s \binom{-j+N+r-2}{s}}{(N-s) (s+1)}
\end{equation}
and compute the recurrence
\begin{multline*}
(-j+N+r-1)F(N,j,r)+\\
(j-r+1) F(N,j,r+1)=\frac{1}{-j+N+r}.
\end{multline*}
Next, we solve the recurrence and find the general solution
$$c\frac{(-j+N-1)_r}{(-j-1)_r}+ \frac{1}{(N+1) (-j+N+r-1)}$$
with the generic constant $c$;
here $(a)_k$ denotes the Pochhammer symbol defined by $(a)_k=a(a+1)\dots(a+k-1)$ for $k\geq1$ and $(a)_0=1$. Finally, we determine the first initial value
\begin{multline*}
F_0(N,j,0)=-\frac{1}{(j-N+1) (N+1)}\\
+\frac{(-1)^N (j+1)!}{(N-1) N (N+1) (2-N)_j}
\end{multline*}
by applying again our summation strategy. This tells us that we have to choose
$c=\frac{(-1)^N (j+1)!}{(N-1) N (N+1) (2-N)_j}$
so that both sides of
\begin{multline*}
F_0(N,j,r)=\frac{1}{(N+1) (-j+N+r-1)}\\
+\frac{(-1)^N (j+1)!(-j+N-1)_r}{(N-1) N (N+1) (-j-1)_r (2-N)_j}
\end{multline*}
agree for $r=0$. As a consequence, the identity holds for all $r\geq0$ for generic values $j,N$. By careful checks it turns out that it holds for all $N,j,r$ with $N\geq2$, $0\leq j\leq N-2$ and $0\leq r\leq j+1$. Summarizing, we transformed the definite hypergeometric sum~\eqref{Equ:InnerSum1} to a linear combination of hypergeometric terms. At this level, we could have used also the well--known hypergeometric machinery presented in~\cite{AequalB}.

Now we turn to the double sum in~\eqref{Equ:TripleSum} which we already simplified to a definite single sum with hypergeometric terms involved:
\begin{multline*}
F_1(N,j)=\sum_{r=0}^{j+1} \frac{(-1)^r \binom{j+1}{r} r!}{(-j+N+r)!} F_0(N,j,r).
\end{multline*}
Again, we compute a recurrence with \SigmaP:
\begin{multline*}
(-j-2) F_1(N,j)-F_1(N,j+1)\\
=
\frac{j-N}{(N+1) (-j+N-2) (N-j)!}\times\\
\times\Big(1
-\frac{(-1)^N (j+2) (j+1)! }{(N-1) N (2-N)_j}\Big).
\end{multline*}
Solving this recurrence, we end up at the following alternative representation
\begin{multline}\label{Equ:F1Simpli}
F_1(N,j)=(-1)^j(j+1)!\times\\
\Big[\frac{1}{N!}\Big(\tfrac{(-1)^N
(j+2)}{(N+1)^2 (-j+N-1)}+\tfrac{N^2+1}{(N-1) N (N+1)^2}\Big)\\
+\frac{1}{N+1} \sum_{i=1}^j \tfrac{(-1)^{i}}{\big(N-i\big)!
\big(i+1\big)! \big(N-i-1\big)}\Big]
\end{multline}
in terms of indefinite nested product--sums in $j$. Now we are ready to deal with the triple 
sum~\eqref{Equ:TripleSum} itself which is given in the form
$$F(N)=\sum_{j=0}^{N-2}  (-j+N-2)! F_1(N,j)$$
where the summand $F_1(N,j)$ is given by~\eqref{Equ:F1Simpli}.
Creative telescoping in the general setting of difference fields leads to the following recurrence
\begin{multline*}
(N+1) F(N)+(-N-2) F(N+1)=\\
=\frac{N^3-4 N-4}{N^2 (N+1) (N+2)^2}\\
+\frac{(-1)^N(N^4+6 N^3+5 N^2-2
N-4)}{(N-1) N^2 (N+1)^2 (N+2)^2}\\
+\frac{(-1)^N N!}{(N+1)(N+2)} \sum_{i=1}^{N-2}
\tfrac{(-1)^i}{(-i+N-1)(i+1)! (-i+N)!}
\end{multline*}
where the right hand side can be simplified further to
\begin{multline*}
(N+1) F_2(N)+(-N-2) F_2(N+1)=+\frac{4 (-1)^N}{N^2 (N+2)^2}\\
+\frac{-N^3-4 N^2-8 N-4}{N^2 \
(N+1)^2 (N+2)^2}+\frac{S_1(N)}{(N+1) (N+2)}
\end{multline*}
applying our strategy again. To this end, we solve the recurrence and find the closed form
\begin{multline}\label{Equ:RunningExp:Res}
F(N)=\frac{-N^2-N-1}{N^2 (N+1)^3}+\frac{(-1)^N \big(N^2+N+1\big)}{N^2 \
(N+1)^3}\\
+\frac{S_1(N)}{(N+1)^2}-\frac{S_2(N)}{N+1}-\frac{2 S_{-2}(N)}{N+1}
\end{multline}
of the triple sum~\eqref{Equ:OrgSumRunningExp}.

\section{Some recent calculations}\label{Sec:Examples}

Subsequently, we present some selective examples from our recent computations~\cite{LoopLegPhysics}. Here several thousand multi--sums have been simplified to indefinite nested product--sum expressions following the strategy from above. In order to carry out these computations mechanically (i.e., by pressing just one button), a new package called \texttt{EvaluateMultiSums} has been developed that combines the different methods of \SigmaP\ and \HarP.

\subsection{All N-Results for 3--Loop Ladder Graphs}

Consider, e.g., the diagram containg three massive fermion propagators,
\begin{equation}\label{Equ:Diag9}
\resizebox{3cm}{!}{~\includegraphics{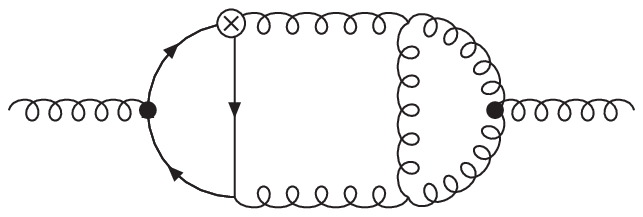}}.
\end{equation}
From its Feynman integral representation around 1000 sums have been produced
automatically~\cite{LoopLegPhysics} whose combination leads to an alternative representation in the Mellin variable $N$. Now the main task is to simplify these sums further. One of the simplest sums is~\eqref{Equ:OrgSumRunningExp} which simplifies with \SigmaP\ to~\eqref{Equ:RunningExp:Res}.
A more typical sum (among the 1000 versions) is
\begin{multline*}
\sum_{j=0}^{N-2}\sum_{s=1}^{j+1}\sum_{r=0}^{N+s-j-2}\sum_{\sigma=0}^{\infty}
 \frac{(-1)^{s+r}S_1(r+2)}{(N
   -r) (r+1)
   (r+2)}\times\\
\tfrac{
   \binom{j+1}{s}\binom{-j+N+s-2}{r}(N-j)!
   (s-1)! \sigma!}{(-j+N+\sigma+1)
   (-j+N+\sigma+2)
   (-j+N+s+\sigma)!}.
\end{multline*}
Following the same mechanism as described in Subsection~\ref{Sec:DetailedExp}, \SigmaP\ simplifies this quadruple sum to indefinite nested sums. Finally, using the \HarP\ package we transform these sums further to an expression involving 145 $S$--sums where the most complicated instances are
$$S_{2,1,1,1}(-1,2,\frac{1}{2},-1;N),S_{2,1,1,1}(1,\frac{1}{2},1,2;N).$$
As it turns out, all 1000 sums can be represented in terms of these generalized harmonic sums. Combining these expressions leads to a rather big expression for diagram~\eqref{Equ:Diag9} containing in total $533$ $S$-sums.

At this point we remark that various relations among the found $S$-sums occur. Now we could use \SigmaP\ again to find a sum representation where all the occurring sums are algebraically independent~\cite{Schneider:10c}. Since this would take quite a while, we use another important feature of \HarP\ following ideas from~\cite{Bluemlein:04}: it contains, e.g., efficient routines and for speed up also precomputed tables that produce representations of harmonic sums~\eqref{Equ:HarmonicSums} (up to weight 8, i.e., $|m_1|+\dots+|m_k|\leq8$) and representations of $S$-sums~\eqref{Equ:SSums} (up to weight 6 where the $x_i$ are in certain relations, like for instance  $x_i\in\{-2,-1,-1/2,1/2,1,2\}$) that are algebraically independent. Using this feature, all $S$-sums cancel, and we end up (within seconds) at a result for diagram~\eqref{Equ:Diag9} where the following sums remain
\begin{align*}
&S_{-4}(N), S_{-3}(N), S_{-2}(N), S_{1}(N), S_2(N), S_3(N),\\
& S_4(N),S_{-3, 1}(N), S_{-2, 1}(N),S_{2, -2}(N), S_{2, 1}(N),\\
& S_{3, 1}(N), S_{-2, 1, 1}(N), S_{2, 1, 1}(N).
\end{align*}

\subsection{3-Loop All N-Results for the $N_f$ Contributions}

E.g., for the diagram
with an outer massive and an inner massless fermion line,
\begin{equation}\label{Equ:FDiag}
\resizebox{3cm}{!}{~\includegraphics{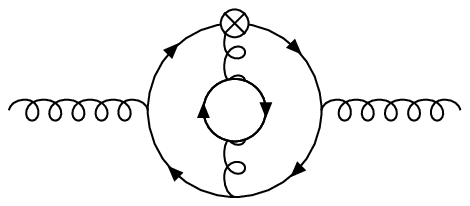}}
\end{equation}
768 multi-sums have been set up which have been simplified to indefinite nested product-sums. One of the simple examples is
\begin{align*}
\sum_{j=1}^{N-2}& \frac{j (j+1) (j+2) (N-j) (j-1)!^2 (-j+N-1)!^2}{-j+N-1}\\
&=\frac{\big(-N^3-5 N^2-4 N+6\big) (N!)^2}{(N-1)^2 N^2}\\
&+\frac{3}{2}\frac{(N!)^2\big(N^3+6 N^2+11 N+6\big)}{(N-1) N
(2 N+1) \binom{2 N}{N}}\sum_{i=1}^N \
\frac{\binom{2 i}{i}}{i}.
\end{align*}
Combining the 768 sums, the final expression for~\eqref{Equ:FDiag} requires 9 MB of memory and is given 
in terms of 703 indefinite nested product--sums which are either harmonic sums, $S$--sums or of the form 
as given in~\eqref{BinomialSums}. To this end, \SigmaP\ finds a sum representation where all sums are 
algebraically independent against each other. In this case, all the unexpected sums cancel, and we get 
the compact form

\allowdisplaybreaks[4]
\footnotesize
\begin{align*}
&\hspace*{-0.3cm}-\tfrac{20 S(1,N)^4}{27 (N+1) (N+2)}+\tfrac{32 \left(6 N^3+61 N^2-21 N+24\right) S_1^3}{81 N^2 (N+1) (N+2)}\\
&-\tfrac{16 \left(48 N^5+746 N^4+2697 N^3+2746 N^2+1104 N+240\right)
   S_1^2}{81 N^2 (N+1)^2 (N+2)^2}\\
&+\tfrac{32}{243 N^2 (N+1)^3 (N+2)^3} \Big(264 N^7+4046 N^6+21591 N^5\\
&+52844 N^4+74856 N^3+66812 N^2+30576 N+2640\Big) S_1\\
&-\tfrac{32}{243 N (N+1)^4 (N+2)^3}\Big(363 N^7+6758 N^6+41285 N^5\\
&+121235 N^4+190235 N^3+150758 N^2+46964 N+2904\Big)\\
& +\Big(-\tfrac{40 S_1^2}{9 (N+1) (N+2)}+\tfrac{32 \left(6 N^3+61 N^2-21 N+24\right) S_1}{27 N^2 (N+1)
   (N+2)}\\
&-\tfrac{16 \left(124 N^5+198 N^4-2387 N^3-6162 N^2-3632 N-480\right)}{81 N^2 (N+1)^2 (N+2)^2}\Big) S_2\\
&   +\left(-\tfrac{32 \left(9 N^3-623 N^2+894 N+276\right)}{81 N^2 (N+1) (N+2)}-\tfrac{160 S_1}{27 (N+1) (N+2)}\right) S_3\\
&-\tfrac{8 \left(56 N^2+169 N+112\right)
   S_4}{9 N (N+1) (N+2)}-\tfrac{4 \left(48
   N^2+101 N+96\right) S_2^2}{9 N (N+1) (N+2)}\\
&+\left(\tfrac{64 S_1}{3 (N+1) (N+2)}-\tfrac{128 \left(N^3+9 N^2-10 N-6\right)}{9 N^2 (N+1) (N+2)}\right)
S_{2,1}\\
&+\tfrac{64 S_{3,1}}{3 (N+1)
   (N+2)}+\tfrac{64 \left(3 N^2+7 N+6\right)}{3 N (N+1) (N+2)}\text{$S_{2,1,1}$}\\
&\hspace*{-0.2cm}+\text{\normalsize$\zeta_2$} \Big(\tfrac{8 S_1(N)^2}{3 (N+1) (N+2)}+\tfrac{16 \left(3
   N^3-N^2+30 N+12\right) S_1}{9 N^2 (N+1) (N+2)}\\
&-\tfrac{16 \left(3 N^3+2 N^2+17 N+6\right)}{9 N (N+1)^2 (N+2)}-\tfrac{8 \left(4 N^2+9 N+8\right)
S_2}{3 N (N+1)
   (N+2)}\Big)+\\
&\hspace*{-0.2cm}+\text{\normalsize$\zeta_3$}
   \left(\tfrac{448}{9 (N+1) (N+2)} -\tfrac{448 S_1}{9 (N+1) (N+2)}\right)
\end{align*}
\normalsize
in terms of the algebraically independent set of harmonic sums
\begin{align*}
S_1(N),S_2(N),S_3(N),S_4(N),S_{2,1}(N),\\
S_{3,1}(N),S_{2,1,1}(N).
\end{align*}

\section{Conclusion}

We demonstrated how \SigmaP\ assists in the task to transform multi--sums to indefinite nested 
product--sum expressions by using the summation paradigms of telescoping, creative telescoping and 
recurrence solving in the setting of \pisiSE-difference fields. It turns out that this mechanism works 
for all instances of our performed 2--loop and 3--loop calculations. Since these product--sum 
representations extend substantially the class of harmonic sums and $S$--sums, these algorithms will be 
useful for further calculations that might leave the solution space of harmonic sums and $S$--sums.

To perform these computations automatically, a new package called \texttt{EvaluateMultiSums} has been developed by the fourth author. During these computations, the package \HarP~\cite{Ablinger:09} based on ideas of~\cite{Vermaseren:99,Bluemlein:99,Moch:02,Bluemlein:04,Blumlein:2009ta,ABS:10} plays an important role to find harmonic sum and $S$--sum representations in optimal form.

\bibliographystyle{h-elsevier2}
\bibliography{citation}

\end{document}